# Mechanisms of Radiation-Induced Degradation of Hybryd Perovskites Based Solar Cells and Ways to Increase Their Radiation Tolerance

*Boris L. Oksengendler[1]\*, Nigora N. Turaeva[2],*
*Marlen I. Akhmedov[3]*

[1]Institute of Ion-plasma and laser technologies, Tashkent, Uzbekistan
[2] Department of Biological Sciences, Webster University, US
[3]National University of Uzbekistan, Tashkent, Uzbekistan

## ABSTRACT

The basic processes of perovskite radiation resistance are discussed for photo- and high-energy electron irradiation. It is shown that ionization of iodine ions and a staged mechanism of elastic scattering (upon intermediate scattering on light ions of an organic molecule) lead to the formation of a recombination center I$i$. The features of ionization degradation of interfaces with both planar and fractal structures are considered. A special type of fractality is identified, and its minimum possible level of photodegradation is predicted. By using the methodology of classical radiation physics, the Hoke effect was also studied, as well as the synergetics of cooperative phenomena in tandem systems. The principal channels for counteracting the radiation degradation of solar cells based on hybrid perovskites have been revealed.

**Keywords**: hybrid perovskite, radiation and degradation of solar cells, tandems, fractal interfaces, Hoke effect, scale-free network

---

\* corresponding author e-mail: oksengendlerbl@yandex.ru

## 1. INTRODUCTION

In the long history of the creation of solar cells (SCs) based on various semiconductor materials, it has become absolutely clear that the issue of radiation stability of these semiconductor structures has always remained one of the most challenging [1-4]. At present, the established views are that the issue of SC radiation degradation has three aspects:

-degradation of an SC under the action of solar light in a wavelength range considered as the operating one;

-degradation of a SC under the action of high-energy radiation;

-degradation of a SC under the conditions of radiation technology;

The first aspect is important for common operating conditions when photochemical processes are in the foreground. The second aspect is vital for operation in space and under special terrestrial conditions connected with intense radiation (atomic power plants, conditions of a nuclear explosion, etc.). The third aspect is the degradation that always takes place immediately in the moment of constructive radiation technology, and it is necessary to establish when radiation technological processes begin to be leveled by radiation degradation processes.

The mechanisms implementing all these three aspects of radiation physics are utterly different, since, different ratios between the four factors of radiation action (ionization, shock wave, heat release, and elastic scattering) occur in each of them [5–7].

The so-called organic-inorganic perovskites, which combine atoms of metals, haloids, and organic molecules in their regular structure [8], were an absolutely particular new type of semiconductor and appeared to be promising at the present stage of the conversion of solar energy into electrical energy. This result refers to both: one transitional and tandem structures [9–12]. After its synthesis, the importance of the problem of SC radiation degradation, which remains no less important (as compared to silicon, gallium arsenide, and cadmium–mercury–tellurium alloys), looks much more complicated already from a priori notions. But what is this connected with?

First, it is connected with the multicomponent composition of organic-inorganic perovskite material, which is such that strong asymmetry of the masses of components takes place.

Second, in this type of substance, completely different types of chemical bonds are implemented.

Third, states with quite a high band-gap energy are implemented.

Fourth, the properties of both inner and outer interfaces begin to play a special role.

Fifth, both ordered and disordered atomic structures are often implemented in these materials.

Such a combination of properties, which resemble systems with high-temperature conductivity to some extent [13], becomes the basis for diversity of radiation physics and chemistry [14]. The latter, in turn, requires detailed knowledge of the micromechanisms of the radiation response of this class of perovskites and the devices based on them in order to use the approaches of radiation technology, as well as a search for ways to reliably control the high radiation stability. The goal of the present work is to describe the approach to the mechanisms of radiation degradation of organic-inorganic perovskites and some device structures based on them in order to formulate the main trends of the future construction of the radiation physics of this class of materials.

## 2.  GENERAL METHODOLOGY OF RADIATION PHYSICS

Radiation physics of solid states [5,7] arose in connection with the problem of the atomic bomb in the middle of the twentieth century. Over the past years, this science has become a deep and rich field of physics, chemistry and biology, which has major achievements and introduced new concepts to all physics. The methodology of modern radiation physics, in addition to classical quantum and statistical physics of condensed matter, is fundamentally connected with such concepts as: "nano", "fractal", "synergetics", and "chirality". At the same time, in studying radiation effects in condensed matter, the use of classical concepts of various channels of radiation energy transfer to the irradiated medium, as well as concepts of elementary radiation-stimulated atomic rearrangements (radiation defect formation, RDF; radiation-stimulated diffusion, RSD; radiation-stimulated quasi-chemical reactions, RSQCR; radiation-stimulated movement of boundaries, RSMB; radiation-stimulated movement of dislocations, RSDM; the radiation stimulated disordering, RSDis). All these elementary atomic rearrangements can be classified as thermal and athermal radiation effects due to their thermal and electron-stimulated origin, and low-scaled or large-scaled radiation effects.

These basic concepts are summarized in two tables (1 and 2). Further, it must be noted that the formation of radiation defects is the primary and most important step of the overall radiation effect. The final resolution of the radiation-physical problem is considered to be the establishment of the mechanism of the radiation effect. Everything said above is reflected in this article, starting with its name and further.

**Table 1. Dominant channels of energy transfer from radiation to materials**

| Radiation→ Excitements ↓ | Light | X-ray | γ-ray | Electron beam | Energetic ions | Fission fragment | Neutron |
|---|---|---|---|---|---|---|---|
| Elastic displacement | | | x | x | x | | x |
| Ionization | x | x | x | x | x | x | |
| Heat releasing | | | | x | x | x | x |
| Elastic waves | | | x | | x | x | x |
| Shock waves | | | | | x | x | x |

**Table 2. Basic elementary processes of atomic reconstructions in condensed matter**

| Energy Transfer channels → Basic atomic processes ↓ | Elastic displacements | Ionisation | Heat release | Elastic waves | Shock waves |
|---|---|---|---|---|---|
| RDF | x | x | x | x | x |
| RSD | x | x | x | x | x |
| RSQCR | | x | x | x | x |
| RSMB | | x | x | | x |
| RSDis | x | | x | | x |

RDF (Radiation defect formation), RSD (Radiation simulated diffusion), RSQCR (radiation simulated quantum chemical), RSMB (Radiation stimulated motion of boundaries), RSDis (Radiation stimulated disordering )

## 3. MECHANISMS OF DEGRADATION UNDER THE IMPACT OF SOLAR RADIATION

A vital feature of organic-inorganic perovskites is that this material features a large fraction of complexity in chemical bonds. In fact, its structural formula looks as $CH_3 - NH_3^+ Pb^{2+} I_3^-$. It is thus evident that the binding energy (potential well) results from the electrostatic interaction manifesting itself in a high Madelung energy value [9, 12]. In particular, the binding energy of an $I^-$ ion in its regular position turns out to be 7.75 eV.

Since the basic effect of photogeneration in perovskite $CH_3 - NH_3^+ Pb^{2+} I_3^-$ is electron transfer from the valence band to the conduction band of the perovskite and these bands are formed from the states of $I^-$ and $Pb^{2+}$, correspondingly, then, from the point of view of crystal chemistry [15], such a photogeneration act converts the negative iodine ion into a neutral atom, which eliminates the Madelung potential well for regular iodine, and its slightest fluctuation leads to displacement into an interstitial position $I_i$. Therefore, the schematic representation of successive photoelectron- ion processes looks like:

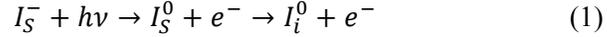

$$I_S^- + h\nu \rightarrow I_S^0 + e^- \rightarrow I_i^0 + e^- \qquad (1)$$

However, implementation of the latter part of the chain (which would always take place in a gas phase) is impeded by the process determined by the quantum nature of a formed hole [16]. In fact, a neutral iodine atom (in the scheme of crystal chemistry) is a hole localized on anincidental iodine. Note that such a duality (crystal chemistry and band approach) is a typical dichotomy popular in the solid-state physics [17].

Next, the initially localized hole is delocalized along the iodine chain; therefore, the characteristic delocalization time can be estimated from the Bohr– Heisenberg uncertainty relation [18]

$$\tau_h \approx \frac{\hbar}{\Delta E_V}, \qquad (2)$$

where $\Delta E_V$ is the valence-band width.

Consequently, an alternative of either translation of a hole and back conversion of $I_S^0$ into $I_S^-$ or displacement of $I_S^0$ into an interstice due to the quantum nature of a hole will be determined by an expression [5]

$$\eta = \exp\left(-\frac{\Delta E_V}{\omega_D h}\right) = \int_{\tau_\dagger = 1/\omega_D}^{\infty} W(t)dt \qquad (3)$$

The described mechanism of defect formation, which constitutes a fundamental act of photodegradation, is essentially a variant of the so-called Dexter– Varley mechanism [21] in radiation physics of ionic crystals [22] in its quantum variant developed 40 years ago by a research team in Tashkent as applied to subthreshold defect production [19, 23, 24]. Let us add that $\omega_D$ is the Debye frequency and $\eta$ is the probability of formation of a defect (in this case, an interstitial $I_i$, which is the main recombination center [11]).

In relation to perovskites, we pointed out the special importance of this degradation mechanism earlier in [25], wherein it was shown that, in the bulk of a crystal, $\eta$ is on the order of $10^{-4}$. A number of new features of this mechanism are expected to be observed at a higher quantum energy than that corresponding to the hump of the solar spectrum, when the initial ionization can proceed in a subvalence shell of perovskite (Figure. 1(a)).

The main channel of decomposition of thus formed "subvalent" hole is a band-to-band Auger process. As a result of an Auger process, a hole "floats" up to the valence shell yet "proliferating" to yield two holes.

The latter leads to still greater Coulomb instability upon the transition from the initial $I_S^-$ into $I_S^+$, but how will it affect the probability of displacement $\eta$?

Studies show [25, 26] that the double charge on $I_S^+$ locally decreases the valence-band width $\Delta E_V$. In addition, the time $\tau_\dagger$ needed for the knock-out of $I_S^+$ ion into an interstice also diminishes (Figure. 1(b)). As a result, the probability of displacement $I_S \rightarrow I_i$ rises considerably; i.e., the fundamental degradation of a SC upon the addition of ultraviolet radiation increases.

The summary of the above is also evident: in all places in a crystal where structural imperfections (locally) decrease $\Delta E_V$, the probability of defect production is increased. It is necessary to add that the existence of the process described by the value $\eta \ll 1$ is evidence that the scheme of photodestruction, according to Schoonman [27], is apparently overestimated in the role of photodegradation. In fact, one may note that, in an experimental work [28], a serious delay in the decomposition of the state immediately following photogeneration of an electron–hole pair is pointed out. However, it is more correct here to speak not of the slowing of this stage but simply of the small fraction of the primary instable state that is transformed into defects, which manifests as a "delay" of the defect-formation act.

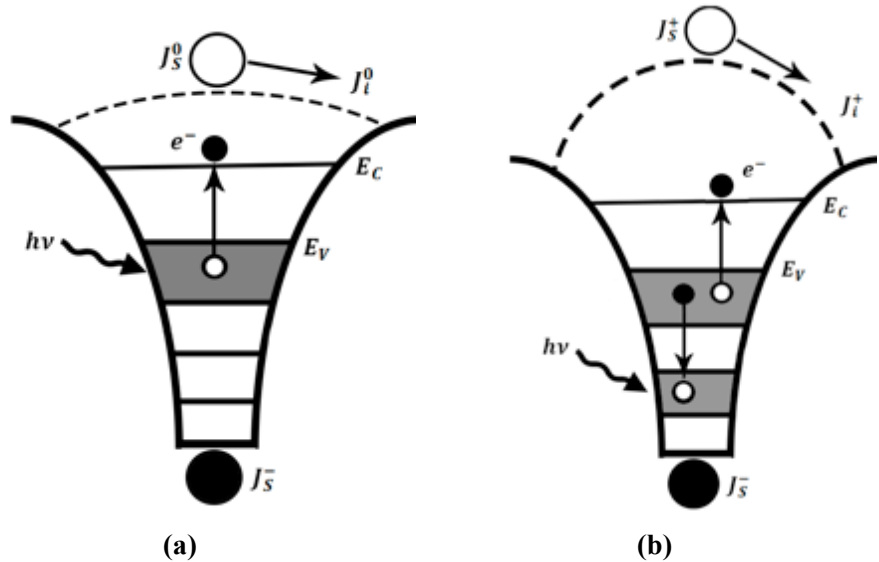

**(a)**                            **(b)**

Figure 1.

(a). Two-stage schematic representation of displacement of ion $I_S^-$ in a lattice position into an interstice ($I_S^- \to I_S^0 \to I_i^0$) upon the absorption of a photon in the valence band.

(b). Three-stage schematic representation of displacement of ion I in a lattice position into an interstice ($I_S^- \to I_S^0 \to I_S^+ \to I_i^+$) upon the absorption of a photon in the "subvalence" shell.

## 4. MECHANISM OF DEGRADATION UNDER THE ACTION OF HIGH-ENERGY PARTICLES (A SPECIAL ROLE OF AN ORGANIC MOLECULE)

Let us discuss the most popular mechanism of defect formation, viz., the displacement of a regular atom upon elastic scattering of a fast extraneous particle on it [5, 6], taking into account the features of the structure of organic-inorganic perovskites. The efficiency of the displacement of a regular atom upon elastic impact in a simplified form is characterized by the defect-formation cross-section [29]:

$$\sigma = \int\limits_{E_d}^{E_{MAX}} d\sigma_T = \pi\sigma_0 \left(\frac{E_{MAX}}{E_d} - 1\right) \tag{4}$$

where $\sigma_0$ is a particular characteristic of Rutherford scattering of a charged extraneous particle (e.g., elec-tron) on an atom, $E_d$ is the characteristic displace-ment energy (effective potential-well depth) of a regular atom in a crystal, and $E_{MAX}$ is the maximum possible energy transfer from a fast extraneous particle to a regular atom in a crystal at head-on collision. It is of significance is :

$$E_{MAX} = 4\frac{m_e M_a}{(m_e + M_a)^2} E_e \tag{5}$$

where $m_e$ is the mass of a fast electron, $M_a$ is the mass of a regular atom, and $E_e$ is the fast electron energy. Combination of equations (4) and (5) indicates the principal channel of increase in the efficiency of defect formation, which is a large $E_{MAX}$ value.

Turning back to organic-inorganic perovskites, it is worth noting that, in such a material composed of atoms with a very great difference in masses of atoms (H and I, Pb), an extraordinarily efficient channel of defect formation arises: upon elastic scattering, fast electrons displace hydrogen atoms ($M_H$), which are next scattered on an iodine atom (with a mass of $M_I$) and displace it into an interstice. That is how the basic recombination center $I_i$ is formed in perovskites [30].

It can be shown readily that the ratio between the maximum energy transfer to a regular iodine atom for the staged and direct processes has the form [5]

$$\frac{E_{MAX}^{(2)}}{E_{MAX}^{(1)}} = \frac{4M_I^2(m_e + M_H)^2}{[(M_I + M_H)^2(m_e + M_I)^2]} \approx 4\left(\frac{M_H}{M_I}\right)^2 \ll 1 \tag{6}$$

Note that the possibility of two-stage defect formation was qualitatively discussed as applied to hydrogen-doped germanium [31]. However, these are the organic-inorganic perovskites for which staged degradation is the basic fundamental degradation process, since, in their case, the concentration of the light component is the same as that of the others (and that is what makes the degradation in perovskites utterly different from this case [31]).

On the other hand, concern about the pronounced ionicity of the crystal components is of importance for the elastic mechanism of defect formation in perovskites. In fact, microscopic theory [5, 32], which makes it possible to express the displacement energy $E_d$ (Eq. (3)) in terms of the depth of a true potential well of a regular atom and its capture radius $R_0$ (Vineyard–Koshkin "instability zone" [5, 29]), indicated a strong increase in $E_d$ for ionic solids as compared to the other types of chemical bonds; therefore, $dE_d/d\alpha_i > 0$ (where $\alpha_i$ is the degree of ionicity in chemical bonds [17]). Therefore, defect formation in perovskites is expected to be problematic $\left(d\sigma_d/d\alpha_i < 0\right)$, which is indicative of the factor of their high native radiation resistance and, to some extent, should compensate for the efficiency of the staged channel of defect formation, turning the action of any energy radiation into internal proton irradiation. It should be mentioned that, in this case, the energy threshold of damage of a perovskite crystal is decreased drastically (Figure 2).

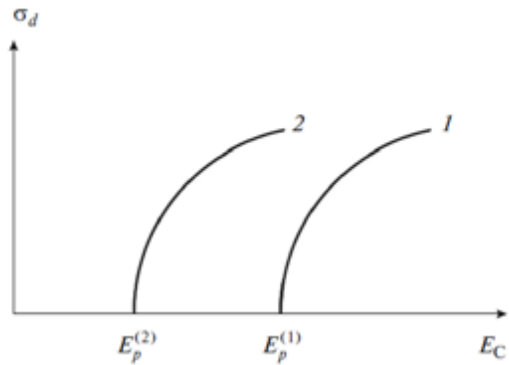

Figure 2. Energy dependence of the cross section (probability) of defect formation upon a (*1*) direct or (*2*) staged process of elastic displacement of a heavy atom (I) into an interstice.

## 5. FEATURES OF RADIATION DEGRADATION OF SCs WITH FRACTAL INTERFACES

An unusual property of SCs based on organic-inorganic perovskites was recently discovered [33]: it turned out to be that, all other things being equal, the transition from a planar interface architecture (perovskite/HTM) to a rough (fractal) one increases the short-circuit photocurrent by 13%. Several reasons for such an extraordinary effect have been proposed [34] (an increase in the total path length of a solar beam in the vicinity of the interface, the facilitated passage of carriers in an extended junction, and the conversion of recombination levels in the interface into the class of attachment levels due to nanofractal fluctuations of the interface potential). Let us dwell on the third affecting channel in greater detail, taking into account both its greatest (probably) significance and rather intriguing demonstration in degradation processes, which has never been discussed before (see below).   Since perovskites are materials with a high degree of ionicity, a Madelung model [35] is quite applicable in light of their electronic states at the interfaces. According to the model, the occurrence of the surface levels (Tamm levels) in the energy gap can be associated with a natural decrease in the Madelung energy at the surface. This idea was extended to the boundary with curvature [36]: it was found that the Tamm surface levels are shifted toward the center of the band gap in convex areas, while the shift occurs closer to the edges of the permitted energy bands in concave areas.

Generalizing this idea, one can say that, by controlling the fractality of the interface, it is possible in principle to transfer the local electronic levels of the boundary from one region to the other, in particular, the recombination levels into the region of attachment levels, for which it is sufficient just to provide the degree of curvature of the concave area such that it intersects the demarcation levels (Figure. 3). We expect that this should decrease the recombination losses of the interface, which forms the basis of explanation of the results in a study [33]. From the viewpoint of radiation destruction of a SC through the modification of interfaces, one more important issue should be noted. In fact, let us consider the Tamm electronic orbitals at the fractal interface (Figure. 4).

As is well known [35], an overlap of neighboring Tamm orbitals creates bands of Tamm states. Evidently, in convex and concave areas of the interface, these overlaps are significantly different: they are smaller in the former and larger in the latter (as compared to a flat interface); the schemes of the valence bands of the corresponding areas agree with this (Figure. 3). This result is fundamentally important for photodegradation (see Section 2); according to its scheme, the probability of defect formation strongly depends on the valence-band width (Eq. (3)):

$$\eta = exp(-const \cdot \Delta E_V) \qquad (7)$$

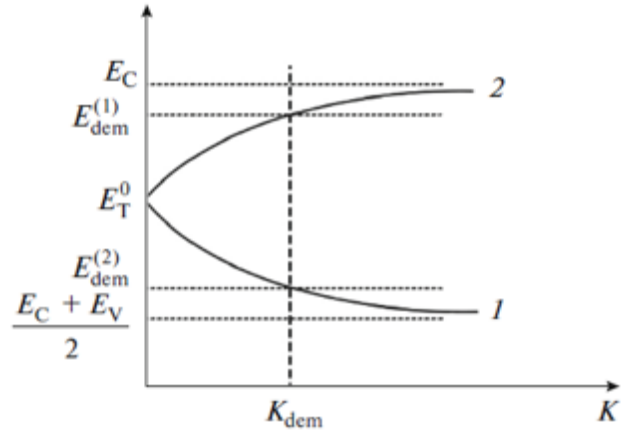

Figure 2. Schematic representation of arrangement of the Tamm surface level with respect to the surface (interface) curvature ( $E_T^0$ is the Tamm level for a flat surface; curves 1 and 2 correspond to convex and concave areas of the surface; $E_{dem}^{(1)}$ and $E_{dem}^{(2)}$ the two possible variants of arrangement of the demarcation level). Upon intersection of the critical degree of curvature $K_{dem}$, the Tamm level passes from the recombination type to the case corresponding to attachment (curve 1) and vice versa (curve 2).

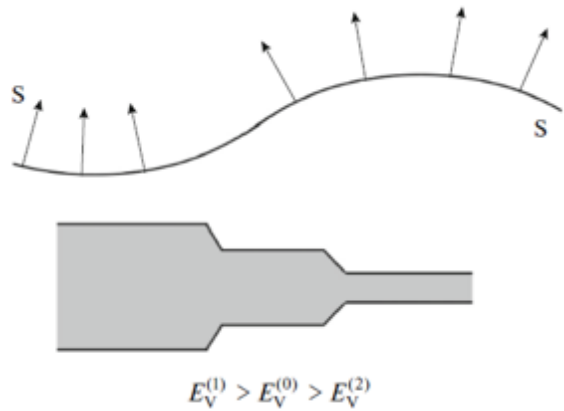

Figure 3. Dependence of the width of the surface band of Tamm states ($E_V^{(i)}$) on the curvature type: (0) intermediate region, (1) concave area, and (2) convex area.

Therefore, in convex areas, the formation of interstitial *Ii* occurs much more intensely than in the concave areas which create diffusion atomic flows from the apices of the relief to the areas of concavities at a fractal surface, i.e., leads to smoothing of the relief and, accordingly, the decrease in the excess photocurrent (the above-mentioned 13% [33]). Finally, on the basis of the above reasoning, one can use interfaces with specially handpicked fractality (Figure. 4), which serves the two noted functions at once: first, it shifts the levels of the interface toward the region of attachment; second, it makes it possible to obtain a low value, that is, radiation resistance (Figure 5). This new idea requires experimental verification. It is worth noting that, generally speaking, smoothing of a fractal surface under the action of radiation has already been observed experimentally in $BaF_2$ , but under strong ionic irradiation [37]. It is thus undoubted that radiation engineering of both exterior and inner surfaces is a promising challenge of the near future [38–40].

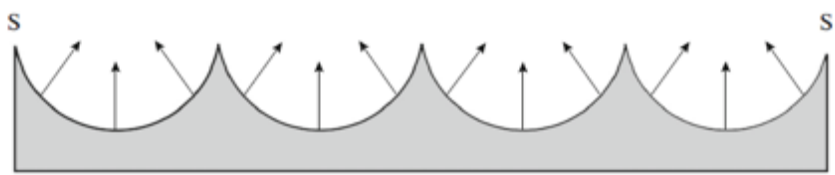

Figure 5. Special form of a rough (fractal) surface (interface) providing minimal photodegradation.

## 6.  RADIATION DEGRADATION AS A RADIATION MACROEFFECT IN PEROVSKITES

In the second and third Sections, only those mechanisms of radiation defect formation that deal with the displacement of regular atoms of a lattice into interstitial positions were discussed. It is necessary to note that the total radiation macroeffect , in this case "radiation degradation" , includes acts of defect formation only as the first stage. The second stage is radiation-stimulated diffusion (RSD) of atoms and defects in the sample, followed by the third stage, quasichemical reactions (QCRs) between atoms and defects, which result in modification of the electronic spectrum (especially in the band gap). If active recombination levels arise in this process, then radiation degradation appears [1 - 7, 29].

An efficient method to describe radiation degradation is the analysis of the kinetics of accumulation of active recombination centers. Such analysis is carried out based on a system of kinetic equations solved under certain initial and boundary conditions. A question arises: what features are imparted to the proceeding of the second and third stages by organic-inorganic perovskite itself and in what way do these features differ from the corresponding stages of the common radiation physics of semiconductors [1–7, 29]? It is significant that this issue is completely unsolved and has even not been posed yet [9–12]. Meanwhile, a range of features of organic-inorganic perovskites, which should appreciably affect the proceeding of the total radiation effect, or degradation, is clear already. Let us discuss them schematically.

### 6.1 Radiation-Stimulated Diffusion

An essential feature of organic-inorganic perovskites is a wide band gap, which allows the electron-stimulated channels of RSD to be implemented upon recharging of defect levels [5–7]. The physics of these acts of RSD is based on the efficient transfer of energy of electron transitions to either deformation of the diffusion potential barrier, a sharp increase in the amplitude of local oscillations of defects, or a burst of temperature in the vicinity of a defect. In the most general case, six modes of intensification of diffusion of defects can be implemented in this process and can be written in the form of the expression for the RSD coefficient [5]:

$$D_{RSD} = D_0 exp\left(-\frac{Q^* - n\hbar\omega_{ph}}{k(T + \Delta T)}\right) \quad, \qquad (8)$$

where $Q^*$, $\Delta E^* = n\hbar\omega_{ph}$ , $T^* = T + \Delta T$ are the diffusion barrier modified with electron–hole junctions, the energy of electron transition, and the burst of temperature, respectively. Of vital importance is the case when $Q^* - \Delta E^* \leq 0$; this is the so-called mechanism of diffusion through inverse recharging [5, 26], a feature of which is the completely athermal character of migration of defects. The experience of radiation physics of common semiconductors [1–7, 29] is evidence that such an effect can be observed with the socalled negative-U centers. There is a high probability that such a negative-U defect is the iodine atom displaced into an interstice; therefore, consecutive iodine recharging is likely to be able to "drive" this defect athermally (the idea that iodine is a negative-U center can be inferred from the features of change in stability of the configuration upon consecutive variation in charge [30]).

### 6.2 Quasi-Chemical Reactions

The high efficiency of defect interaction in perovskites, which leads to the creation of complexes, can be due to two reasons. First, since this material enjoys a considerable degree of ionicity, the opposite charges of defects entail their strong Coulomb interaction (which sharply increases the efficiency of attraction of defects to each other). Second, the great strength of electron–phonon interaction should be taken into account; it determines electron-stimulated atomic rearrangements in a local region [9–12]. In these electron-stimulated processes, a special role may be played by dipole molecules $CH_3NH_3^+$. In fact, any excitation of the regular component of the $CH_3NH_3^+$ lattice will immediately affect the value of the local band-gap energy, which will be modulated when oscillations of separate modes of a $CH_3NH_3^+$ molecule occur (because the Madelung energy in an ionic crystal is an essential component of the band-gap energy) [35, 36]. Oscillations of a $CH_3NH_3^+$ molecule will immediately modulate the Madelung energy and, accordingly, the band-gap energy. It is also of importance that disturbances of the shape of the inner and outer interfaces (in particular, their transition from planar to rough geometry) dramatically change the efficiency of defect sinks. All of the mentioned circumstances will be reflected in change of the coefficients in the corresponding terms of the kinetic equations and, under strongly nonequilibrium conditions, may yield even a synergistic behavior of a crystal [4–6].

### 6.3 Radiation Annealing

An important feature of radiation action on materials is radiation annealing, the essence of which lies in elimination of the radiation effect and, accordingly, radiation degradation. On the basis of the pool of experience of the radiation physics of semiconductors [5], two classes of the processes of this kind should be differentiated: the effect of recombination of electrons and holes on a Frenkel pair $V_I + I_i$ with the subsequent reaction $V_I + I_i \to I_S$, and a special type of quasichemical reactions during irradiation, which is "a reaction of displacement" of an atom $Cl_s$ positioned regularly in the lattice site by interstitial $I_i$ (the case in point is mixed perovskite $CH_3NH_3Pb_{3-x}Cl_x$). This reaction can be written as $Cl_s + I_i \to I_s + Cl_i$ with $Cl_i \uparrow$ denoting rapid diffusion of the displaced atom towards various sinks (interfaces and surfaces). Note that the reaction of a displacement type plays a fundamental role in the radiation physics of all familiar semiconductor materials and is due to vibron effects [5]. Especially notable is the choice of the method of theoretical analysis of the potential relief for both types of reactions. The most adequate approach may be quantum chemistry on the basis of the DFT method [41]. In the Hohenberg–Kohn–Sham formulation, the Kohn–Sham equations reduce a many-particle problem of interacting electrons to a single-particle problem with an effective potential.

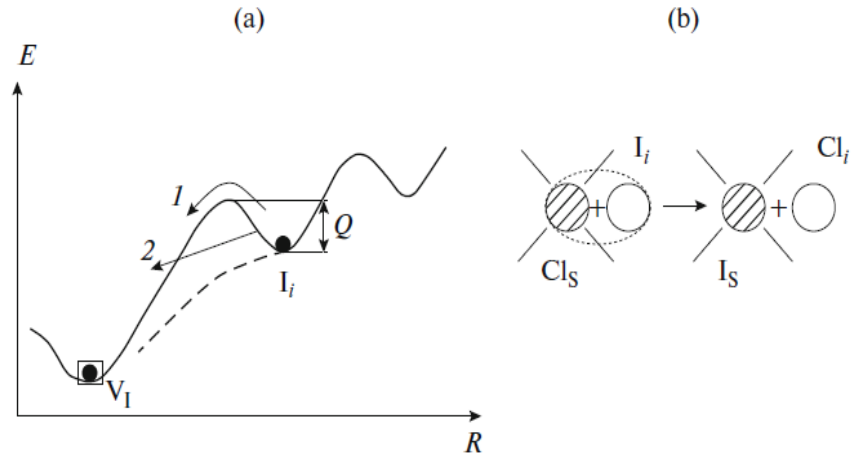

Figure 6. Schematic representation of rearrangement of defects compensating degradation processes: (a) channels of annealing of a close Frenkel pair in sublattice I: (1) thermoactivation channel and (2) recombination-stimulated channel; and (b) "dumbbell" reaction of displacement of regular atom $Cl_s$ by defect

At present, this method is one of the most universal (ab initio) methods for calculation of the electronic structure and various characteristics of many-particlem systems, which is applied in quantum chemistry and solid-state physics. In this method, the many-electron system is described not by a wave function, which would determine a very high dimensionality of the problem equal to at least 3N (the number of coordinates of N particles), but an electron-density function (the function of only three spatial coordinates), which leads to considerable simplification of the problem. In this case it turns out that the major properties of the system of interacting particles can be expressed with an electron-density functional; in particular, according to the Hohenberg–Kohn theorems, which provide a theoretical basis for the DFT method, such a functional is the ground-state energy of the system. As a result, a full picture of electron-stimulated quasi-chemical reactions (Figures. 6(a) and 6(b) can be constructed. Comprehensive knowledge of the potential relief of the reactions, together with the accompanying change in the spectrum of electronic levels in the band gap, allows one to choose the method of striving for an increase in the service life of a device. For example, the regular injection of an electron–hole plasma to the perovskite area can provide systematic purification from close Frenkel pairs (their recombination), according to Figure. 6(a) (arrow 2), similarly to that established in GaAs structures long ago [5].

## 7.   THE HOKE EFFECT IN HYBRID PEROVSKITES

Perovskite solar cells are a champion among other solar cells in the rate at which they increased their power conversion efficiency from 3% to 21% during 7-8 years [42-47]. Ongoing efforts to improve the efficiency beyond the Shockley-Queisser limit are focused on tuning their current bandgap around 1.5-1.6 eV to the range of 1.7-1.8 eV, which is ideal for a perovskite-Si tandem solar cell. This can be achieved by replacing iodide components of $MAlPbI_3$ perovskite with bromide [48-51]. However, this increase does not result in a corresponding increase in open-circuit voltage [52,53], which would be expected due to the increasing of the bandgap. Hoke et al. [53] showed that mixed I/Br perovskite fail to perform because of halide phase segregation induced by light, which is known as the Hoke effect. Solar cells made with mixed halide perovskites containing more than 20% bromide ($MAlPb(Br_xI_{1-x})_3$, $x > 0.2$) showed a decrease in open-circuit voltage with increasing bromide content [52]. It was shown [53] that for materials with $x > 0.2$, the initial PL intensity decreased and a new lower energy peak (as the initial PL peak energy of the $x = 0.2$) is appeared. From these data they assumed that upon illumination two separate phases corresponding to pure bromide and iodide perovskite crystals were developed. Due to the long lifetime and diffusion length of electrons and holes [47,55], they cover multiple grains before the recombination. The thermalization of photogenerated carriers and their further trapping into I-rich low bandgap regions takes place on the picosecond time scale. With illumination, X-Ray diffraction showed the formation of phases with a larger and a smaller lattice constant [53] while without single-phase state. One of the important experimental results is the observation of long-range migration of halides, which could be linked to J-V hysteresis in solar cells and was observed in other halide perovskites [55]. PL and absorption spectra demonstrated also the reversibility of the segregation before and after light irradiation. Bischak et al. confirmed that segregation of I ions take place at grain boundaries and estimated the size of the iodide-rich clusters to be 8-10nm in diameter [55]. Hoke et al. found out that the segregation occurred even at low temperatures [53], but for a longer time period. Bischak et al. also showed that mixed I/Br perovskites with well-mixed structure at room temperature undergo demixing transitions as a function of temperature with a critical temperature of 190K [55]. The phase segregation taking in the wide range of temperature follows Arrenius law with the activation energy similar to halide conductivity activation energies for other halide perovskites [53, 55].

To explain this effect a few theoretical models have been developed [54-58]. Brivio et al. [54] used density functional theory (DFT) to show the thermodynamic characteristics of the solid $MAIPb(Br_xI_{1-x})_3$. Their theory showed that the phase segregation into thermodynamically stable I-rich and Br-rich phases occurred at room temperature upon the provision of necessary energy by illumination. However, the reversibility observed in experiments could not be explained in the frame of the theory. In the comprehensive Bischak model based on molecular dynamic simulations, polarons occur when the free electron and hole generated by light deform the surrounding lattice through electron-phonon coupling. Those polarons funnel into the reduced-band-gap I-rich domains [56]. Their study shows that the unique combination of mobile halides, substantial electron-phonon coupling, and long-lived charge carriers is required for photo-induced phase separation. The authors suggest that the concentrated polaron density switches the shape of free energy from one with one minimum to one with two minima, corresponding to the dark and light states. In this work, we present a mathematical description of the effect based on the Ising model, which was developed for phase transitions in the framework of statistical mechanics approaches.

### 7.1 The Ising Model

The Ising model enabled a mathematical description of phase transformations in ferromagnetics and antiferromagnetics, the order-disorder transitions in binary alloys and lattice gases [58]. In binary alloys AB there are two species A and B, each lattice site can be occupied by either species. Mapping the alloy onto the Ising model can be made by assigning A species to "up" spin $(\sigma_1 = +1)$ and B species to the "down" spin$(\sigma_1 = -1)$. The Hamiltonian of the Ising model for binary alloys can be described by the general formula

$$E = -\sum_{ll'} J_{ll'} \sigma_l \sigma_{l'} - \mu H \sum_l \sigma_l \qquad (9)$$

For the ferromagnetic, μ is the magnetic moment, J is the exchange integral; H is the external magnetic field. By calculating the number of species A and B and the bonds between them, the total energy takes the following expression [58]:

$$E = -\frac{1}{2} zNJ + 2N_{AB}J - (2N_A - N)\mu H \qquad (10)$$

For binary alloys, we have

$$J = \frac{1}{2}\left(\varphi_{AB} - \frac{1}{2}\varphi_{AA} - \frac{1}{2}\varphi_{BB}\right) \qquad (11)$$

$$\mu H = \frac{1}{4}z(\varphi_{BB} - \varphi_{AA}) \qquad (12)$$

Here interaction energies between species such as AA, AB и BB are denoted by $\varphi_{AA}, \varphi_{AB}, \varphi_{BB}$. The exchange integral $J$ of the Ising model plays the role of mixing energy, which gives the tendency for the system to keep different species apart $(J > 0)$ or together $(J < 0)$ giving a uniform mixture. In such binary alloys there is a critical temperature, above of which the lattice is disordered. Below the critical temperature the alloy tends to be ordered either by separation into the domains rich by A or B species$(J > 0)$, or with sites of one of sublattice mostly occupied by a given species in a regular arrangement$(J < 0)$. Note that at low temperatures the ferromagnetic Ising model $(J > 0)$ gives phase separation into +1 – rich and -1 –rich domains, while in the antiferromagnetic Ising model $(J < 0)$ corresponds to antiferromagnetic ordering. By minimization of free energy the relation between the long-range order parameter and the temperature is estimated as

$$\left(\frac{N_A - N_B}{N}\right) = th\left\{\frac{1}{kT}\left(\mu H + \left(\frac{N_A - N_B}{N}\right)zJ\right)\right\} \qquad (13)$$

When the order parameter is equal to ±1, A and B species form separate domains in the lattice. When it is zero, A and B species are mixed in random and there is no long-range order.

We can apply the Ising model developed for binary alloys to the Hoke effect in perovskites described above. The lattice of the Ising model is then related to the lattice, the sites of which are occupied by halide ions. As species A and B introduced in the Ising model we imply two halide anions, iodide and bromide, respectively. In dark conditions, at room temperatures above the critical temperature (190K), the hybrid perovskites can be presented as a lattice with compositional disorder (sites occupied with A or B more and less at random). Under the light, the lattice transits into more ordered state with the rich in A and B phase separate domains, shifting the critical temperature up to higher temperatures. We can account it by adding the term $\delta$ responsible for the interaction of electron-hole pair generated by light with the iodide ions fluctuations discussed above into the formula (13):

$$X = th\left(\frac{1}{kT}(\delta + \mu H + zJX)\right) \qquad (14)$$

Here $X = \frac{N_A - N_B}{N}$ , $\delta = z(\varphi_{AB} - (\varphi_{AA} - \Delta\varphi_{AA}))$, $\Delta\varphi_{AA}$ is the interaction of iodide ions with an electron-hole pair. Since interaction between iodide ions fluctuations and electron (hole) polarons is attractive, and they interact with each other stronger compared to with bromide ions, we have $\delta > 0$, the curve $X(T)$ shifts to the right of room temperatures, to make segregation occurred at room temperatures upon illumination as observed in the experiments [12] (Figure. 7).

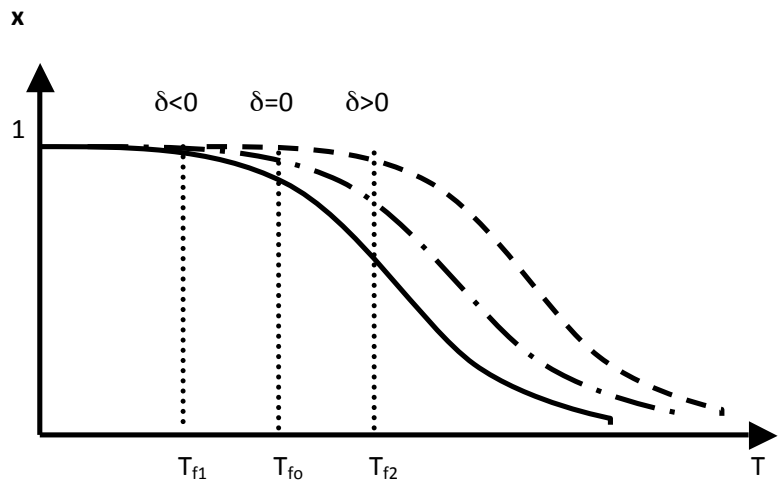

Figure 7. The dependence of ordering at different signs of $\delta$.

### 7.2 The size of ordered domains

It was suggested [55] that the extension of segregated halide atoms rich domains depends upon the extension of the polarons and is in the range of 8-10 nm. Using those data we can evaluate the correlation length of iodide-rich (or bromide rich) domains. The correlation function at large distances $R$ is defined by the formula

$$\Gamma(R) \propto R^{-1}e^{-R/\xi} \tag{15}$$

The size of the domains $L$ is defined from the expression

$$L^2 = \frac{\int R^2 \Gamma(R) d^3 R}{\int \Gamma(R) d^3 R} \qquad (16)$$

Taking into account the estimated value of $L$, we have that    is about several nm

### 7.3 Synergetics of perovskite photoseggregation: intermittency mode

Among the results on photo-segregation, it is very interesting to note some quasi-transitivity in the change in photoluminescence of mixed perovskites $MAIPbI_xBr_y$ [59]. This poses the problem, at least in principle, to modelly understand this quasi-periodicity within the framework of the phenomenology used. Looking at some kind of synergetics in this phenomenon, we will represent the evolution of the phenomenon of photo-aggregation in the form of a one-dimensional mapping for the long-range order parameter

$$X_{n+1} = \varphi(\mu, X_n) \qquad (17)$$

where the index $n$ divides the entire time scale into equal intervals, $\mu$ is the set of parameters of the problem. Using a certain renormalization, expression (9) can be rewritten in the form

$$X_{n+1} = \tilde{\varepsilon} + X_n + u X_n^2 - g X_n^3, \qquad (18)$$

which easily can change to the differential equation:

$$\frac{dX}{d\tau} = \tilde{\varepsilon} + u X^2 - g X^3 \qquad (19)$$

where $\tau$ is the time in a certain normalization, $u, g$ are parameters that take into account both the internal arrangement of the system $(u)$ and its response to the external photoeffect $(g)$. This formulation of the question allows us to successfully apply the laws of nonequilibrium statistical mechanics, in particular to the entrance to the range of variation of the quantity $\to X_{in}$. This allows us to immediately implement the ideas of synergetics, for example using the Lamerey diagram (Figure. 8).

This diagram shows that after entering  the ordering process ($X \to X_{in}$ is the left point on the abscissa axis), the order parameter $X$ slowly changes, which then goes into a quasi-oscillatory turbulent regime with large $X$ drops on each jump. This Lamerey diagram corresponds to the kinetics of intermittency, where laminar and turbulent stages coexist (Figure. 9).

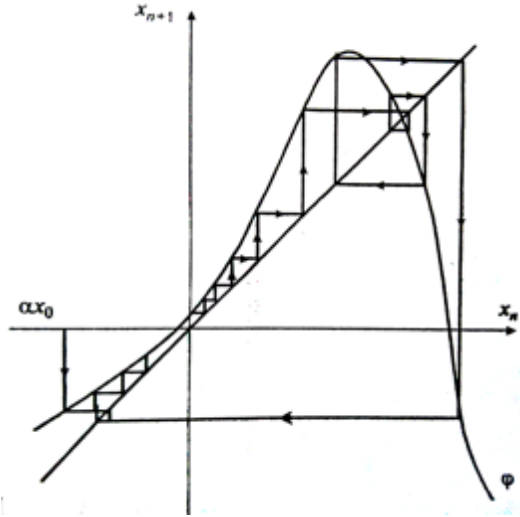

Figure 8. The Lamereya diagram, describing the mode of intermittency of the variation of the long-range order parameter in the photoseggregation of mixed perovskite.

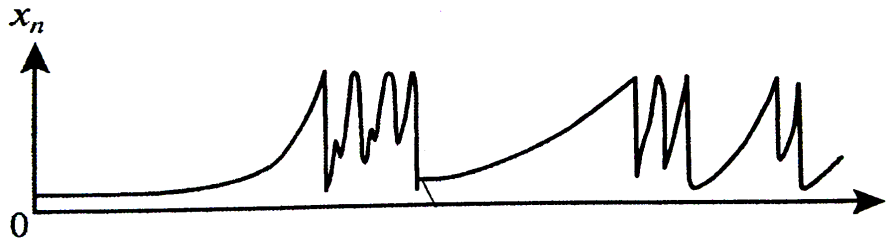

Figure 9. Kinetics of the order parameter for photosegregation of mixed perovskite (conjugation of laminar and turbulent phases of photoluminescence takes place).

Equations (18) and (19) easily allow us to estimate the average time of the laminar regime of an increase in the order parameter $\{X\}$ [60, 61]:

$$\langle \tau \rangle = \frac{1}{\sqrt{\bar{\varepsilon}u}} arctg \left[ \frac{c}{\sqrt{\bar{\varepsilon}u}} \right]. \qquad (20)$$

Here $c \approx X_{max}$ (Figure. 8). Note that for $\frac{c}{\sqrt{\bar{\varepsilon}}u} \to 1$ we have:

$$\langle \tau \rangle \approx \frac{\pi}{2\sqrt{\bar{\varepsilon}}u}. \tag{21}$$

Let us now discuss the "physical content" of the mathematics used. Note that the quadratic term $X_n^2$ in the formula (11) corresponds to such a picture: if there is a spherical region with an intermediate $X_n$, then on the surface of the sphere (that is, on the boundary between ordered and unordered) the value of the contacts is proportional to $X_n^2$ ( this is reasonable for $uX_n^2 > gX_n^3$). On the other hand, when $X_n > {u}/{g}$, the role of photogenerated e-h pairs plays the forefront, the interaction of which with ordered regions (which is modeled by $g$) destabilizes the ordered regions; And this effect is stronger the more $X_n$.

Thus, the developed synergetic model, corresponding to a certain extent and experiment [59], can point to dynamic chaos in the form of an intermittency regime, which is extremely trivial for a given perovskite system and to some extent supplements the number of examples obtained in other areas (for example, the Belousov-Zhabotinsky reaction [61]).

In conclusion, the Ising model of segregation of halide ions upon illumination in hybrid perovskites allows establishing the relationship between the long-range order and temperature describing the transition between demixed and mixed states with taking into account the interaction of iodide ions fluctuations with an electron-hole pair generated by light. The synergetic theory of segregation is also discussed.

## 8.   PROBLEMS OF RADIATION DEGRADATION OF TANDEMS

Along with the development of the problem of radiation degradation of single junction solar cells, on the basis of hybrid perovskites, the complex of questions of tandems (tandem of type 2T with CZTSSe / PVSK structure, 4% conversion efficiency) was also studied, beginning from 2012. To date, both 2T and 4T tandems have been studied on very diverse systems: PVSK / PVSK; organics / PVSK; organics / PVSK (bilayer system); Silicon / PVSK. The structures of the last type are the most promising giving the efficiency of 27%. In principle, tandems were designed to solve several problems. Firstly, spectral losses can be minimized by using two consecutive components with different electronic band gaps ($E_g^{(1)}; E_g^{(2)}$). Secondly, there is a possibility of selecting the composition of components to create so-called "Contour maps". Thirdly, it is also possible to choose the electrical connection between the components of the tandem (2T or 4T). As a result of these possibilities, according to De Vos [64] the theoretical limit of efficiency can be sharply increased up to 47%, in contrast, to 33% by Shockley – Queisser[65] . Against this background, the problems of stability and radiation degradation for tandems are still no less acute. The problem of stability and degradation of tandems has required a broader research than a single-junction structure. Indeed, one should point out the most important aspects:

- (radiation) degradation of each component of the tandem separately;

- (radiation) degradation of the interfaces between the components;

- the problem of choosing the optimal symbiosis of a heterogeneous structure topology and electrical connection between the most important nodes of this structure; the latter should provide special stability to external influences of the common device;

- development of special methods of radiation annealing of defects at the moment of their inception (in situ).

Some new aspects of the problem of stability of solar cells based on hybrid perovskites are discussed below.

### 8.1 Features associated with the intermediate layer

Types of intermediate layers can be different upon specific tasks. An important type of degradation occurs when the role of the intermediate layer is reduced to the object where carrier recombination occurs to provide the general neutrality.

As described above, the most effective neutralization (through carrier recombination) can be accomplished at the interfaces by selecting their fractal properties. This is accomplished by translating the sticking levels to the recombination zone. However, radiation-stimulated diffusion turns the fractal interface into the flat interface, which makes neutralization difficult and takes the device out of the desired mode of operation.

### 8.2 The processes of degradation of the electrical connections of the device

The capabilities of various architectures and combinations of the types 2T and 4T allow identifying the parameters, characterizing the device as a whole. This leads to a consideration of some analogies of a complex ecosystem with external influences. As it is well-known, the most common analysis of the current type of ecosystem sustainability is based on the network theory. In this theoretical construction, several characteristics appear that allow using the topology to describe the entire system. The most important concepts were the so-called clustering, "close world" and scale-free system. Focusing on our specific problem of degradation (in particular, radiation) of the complex device, we highlight the fact that the scale-free character of all elements and connections between them has a unique property of preserving the functionality of the entire system as long as possible, despite the fact that a large ratio of elements is out of operation. Moreover, the description of the damage threshold with the help of percolation ideas turned out to be adequate, and the percolation threshold of scale-free systems turned out to be very high. Applying to the object under study the following technology strategy can be offered - to build the architecture of electronic communication and the individual parts of the device in such a way to get a scale-free network.

It is interesting to note that the problem of resistance to external influences for the type of perovskite instruments under study can also be described in the framework of a simpler and generally accepted theory of ecosystems. This approach uses the analogy of ecosystem participants with electrons and holes of two parts of a common tandem. It was possible to trace the analogy for all three types of relationships between ecosystem participants (the predator – prey model, the symbiosis model, the restricted food model) with the electron – ionic processes of the tandem. A characteristic feature of this approach has become a large range of already known models of ecosystems [69], which gives a hope for the usefulness of this direction.

It is especially interesting to point out the possibility of using the results of game theory [70]. When a tandem is modeled by the interaction of two hierarchical structures (Figure 10), each of the components of the tandem is a hierarchical (two-step) system ($w_1$ and $Q_1$; $w_2$ and $Q_2$). Events at each level of both subsystems are described by stochastic differential equations according to the methods of (Itoh, Stratonovich)[71]. From the Figure 10 it is clear that there are several types of interaction (arrows): the upper levels with the lower ones, and with "their own" and aliens; the latter type of interaction is called cross-correlation. The whole analysis is based on the fact that a certain characteristic (F) of the whole tandem is selected, which is constructed from the partial characteristics of the components ($\Phi_1$ and $\Phi_2$). There are two variants of these constructions: multiplicative ($\Phi = \Phi_1 * \Phi_2$) and additive ($\Phi = \Phi_1 + \Phi_2$). Further optimization occurs which either maximizes $\Phi$ (useful property) or minimizes $\Phi$ (when it is a harmful property). This approach allows us to study the role of various noises and has very rich results from the game theory (see [72]).

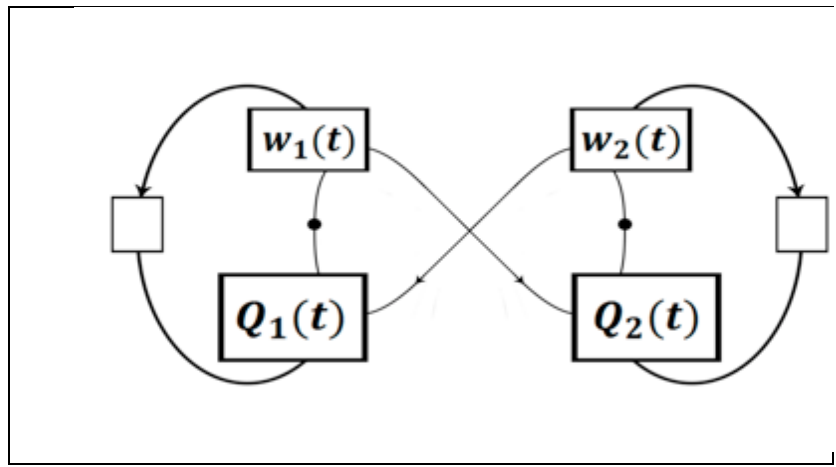

Figure 10. Model of state variables of two communicating hierarchical systems for two components of a tandem based on perovskite, empty squares mean controllers, black dots mean the probability of systems 1 and 2 is in the homeostatic region at Q levels with a corresponding value of the correlation coefficient.

## 9. SUMMARY OF RESULTS ON METHODS TO PREVENT RADIATION DEGRADATION OF PEROVSKITES

We can distinguish two aspects of control radiation degradation:

1) control of the local electronic spectrum in the forbidden zone;

2) the management of geometry and electronic states on the interfaces;

In general, these two aspects are implemented using the following methods:

1) injection annealing (recharging the level of defects);

2) radiation shaking (the liquidation of metastable states by an elastic wave);

3) the fractality of interfaces (the modulation of the width of the allowed zone of valence states);

4) the fractality of the crystal surface (modulation of the width of the allowed zones of Tamm levels);

5) acceleration of quasi-chemical reactions (the realization of reactions between defects through the use of electronic excitations);

6) the possibility of a wide choice of the type of halogen (the effect of changing the width of the allowed valence band affecting the lifetime of the resulting hole);

7) the use of energy laws of game theory to obtain the optimal mode of tandems (the interaction of the hierarchical structures of the two components of the tandem allows using the methods of stochastic differential equations to choose the optimal mode for both positive and negative properties of the structure);

8) the organization of the complex structure of the device in the form of no large-scale networks (the essence of the effect is that without large-scale networks they are particularly resistant to destructive effects, as described by percolation interactions);

## 10. CONCLUSIONS

Radiation degradation of organic-inorganic perovskites and device structures based on them, like any other physical radiation macroeffect, is a combination of three consecutive stages: radiation defect formation, radiation-stimulated diffusion, and quasi-chemical reactions between defects. The implementation of these consecutive stages results in two effects: the electronic spectrum in the band gap and geometry of interfaces change. Both of these effects can dramatically affect the photoelectric parameters of perovskite materials and devices.

Among the processes of defect formation, the most essential are light ionization of the valence and subvalence bands constructed of the orbitals of the iodine components and elastic displacement of iodine anions through the two-stage channel of energy transfer from a fast particle to lattice atoms. For radiation-stimulated diffusion, the most important channels of stimulated displacement of defects are various kinds of recharging of deep-level centers. The effect of light induced smoothing of rough (fractal) interfaces can be due to modulation of the valence-band width by the change in overlap of Tamm orbitals; this smoothing eliminates the increment in the short-circuit photocurrent discovered for rough interfaces.

It is necessary to point out the results of the research of tandems. The degradation properties depend both on the layer between the components and on the properties of the components themselves. Several original methods for analyzing the radiation degradation properties of tandems based on the game theory approach between two hierarchical structures - components of a tandem (Itoh, Stratonovich stochastic differential equations in the regime of general optimization of tandem properties) and the analysis within the model of scale-free networks have been proposed  On the basis of the latter method, it was possible to identify the type of networks modeling tandem structures that are particularly resistive to their degradation.

We note that, in principle, the unique complexity of perovskite instruments and various types of radiation makes it possible to build a "complexity matrix". Horizontally, in such matrix of complexity, properties of perovskites are located, and vertically, various factors of radiation are positioned. The number of elements in such a complexity matrix will indicate the actual possibility of numerous radiation modifications of the device, but at the same time, unfortunately, too many channels of their degradation. With such a unique object, radiation physics encounters for the first time.

Let's take off the hat, gentlemen as
And now remains
That we find out the cause of this effect;
Or rather say, the cause of this defect,
For this effect defective comes by cause:
Thus it remains, and the remainder thus.

(Shakespeare, Hamlet)

## ACKNOWLEDGMENTS

In the course of the research in the field of radiation degradation, the authors had the opportunity to discuss many issues with a number of prominent scientists in the field of condensed matter physics. Among them, we would especially like to thank Academician A.F. Andreyev (Russia), professors: A.A. Zakhidov (USA), N.R. Ashurov (Uzbekistan), Kh.B. Ashurov (Uzbekistan), D.U. Matrasulov (Uzbekistan), Dr. S.E. Maximov (Uzbekistan), Dr. D.R. Aristov (Russia), Dr. N.N. Nikiforova for support and useful discussion of key points. In addition, we would like to express our great gratitude to the students of the National University of Uzbekistan Akramova R.B. and Kobiljonov M.O. for assistance in the design of the chapter.